# Stratification in Randomised Clinical Trials for Rare Diseases and Analysis of Covariance: Some Simple Theory and Recommendations.

Stephen Senn[1,2,3], Franz König[1], Martin Posch[1]

1 Medical University of Vienna, Center for Medical Data Science, Institute of Medical Statistics

2 University of Sheffield, Sheffield Centre for Health and Related Research

3 University of St Andrews, School of Mathematics and Statistics

Correspondence, Martin Posch, Medical University of Vienna, Center for Medical Data Science, Institute of Medical Statistics, Spitalgasse 23, 1090 Wien, Austria, Email: martin.posch@meduniwien.ac.at

A simple device for balancing for a continuous covariate in clinical trials is to stratify by whether the covariate is above or below some target value, typically the predicted median. This raises an issue as to which model should be used for modelling the effect of treatment on the outcome variable, $Y$. Should one fit, the stratum indicator, $S$, the continuous covariate, $X$, both or neither?

When a covariate is added to a linear model there are three consequences for inference: 1) the mean square error effect, 2) the variance inflation factor and 3) second order precision. We consider that it is valuable to consider these three factors separately, even if, ultimately, it is their joint effect that matters.

We present some simple theory, concentrating in particular on the variance inflation factor, that may be used to guide trialists in their choice of model. We also consider the case where the precise form of the relationship between the outcome and the covariate is not known. We conclude by recommending that the continuous covariate should always be in the model but that, depending on circumstances, there may be some justification in fitting the stratum indicator also.

# 1 Introduction

In what follows we shall consider small two-arm parallel group trials in which the target outcome variable is continuous and may be assumed (perhaps after suitable transformation) to be approximately Normally distributed and it is proposed to use a single covariate in analysis. Although the theory we develop applies to trials of any size. its practical importance is greatest for small trials. Thus, it may be helpful for the reader to assume that the context is rare diseases. We shall consider the case of one covariate only and for which the treatment-by-covariate interaction is not fitted However, in the discussion we shall consider briefly what happens if trials become large, if they have more than two arms, if more than one covariate is used or if the interaction is fitted.

A clarification of terminology may be helpful. The term *stratification* usually refers to stratified allocation (sometimes referred to as *pre-stratification*), which is typically accompanied by an analysis taking strata into account. However, even where such stratification has not taken place, it is possible to create strata for analysis. This technique is sometimes referred to as *post-stratification (*also known as *stratification after selection*[1]*)*. In this paper, we are mainly concerned with the former form of stratification, which is to say, trials that have been stratified by design, and unless we explain otherwise, this is the sense that should be understood.

Furthermore, we shall assume that the following specific approach to stratification has been applied to a single continuous covariate, for example the value measured at baseline of the target outcome variable. Alternatively, some sort of prognostic score or 'supercovariate' might be considered[2]. A predicted median value of the variable is used and when the patients are entered onto the trial they are assigned either to the low or high stratum depending on how their value compares to the median. In allocating patients to treatment, an attempt is made to balance numbers on each treatment arm within each stratum. In practice, this would have to be achieved using some such method as permuted blocks within strata[3]. In developing the theory, we shall assume that the blocks are merely a balancing device and can otherwise be ignored. We shall pick this point up in the discussion.

Qu has suggested the following possible classification of models[4] (p233) :

$$
\begin{aligned}
&Model(A): Y \sim Z \\
&Model(B): Y \sim Z + X \\
&Model(C): Y \sim Z + S \\
&Model(D): Y \sim Z + X + S.
\end{aligned}
$$

Here the continuous covariate is $X$ , the outcome variable is $Y$ , the treatment indicator is $Z$ and the stratum indicator is $S$ . We shall adopt this classification system and mainly consider the effect of stratification when it has been decided to fit Model (B). We shall then consider the effect of fitting Model (D).

Note that all four models are main-effect models only. They do not include a treatment-by-covariate interaction. One could consider an expanded set of models in which the interaction of $Z$ with $S, X$ or both was considered. In this paper we shall not cover such models but they will be briefly considered in the discussion. However, as regards stratification there is one important technical point. The trials we are considering are stratified by design. In stratifying, there is a common choice in analysis between either fitting a factor with as many levels as strata in a linear model or fitting the treatment effect stratum-by-stratum and then combining the estimates. The latter also implicitly removes the treatment-by-stratum interaction from the mean square error and raises issues as to how the stratum estimates should be combined, a matter to which the long-standing debate regarding type II and type III sums of squares is relevant[5, 6]. Considering this latter issue would detract from the main points we wish to make, so we shall avoid it by not considering interactions. However, we shall pick this matter up briefly in our conclusions.

The outline of this paper is as follows. In section 2 we introduce the framework we propose to use for discussing the effects of fitting covariates. Sections 3 and 4 cover two aspects that depend on the model employed but are largely independent of the design: the expected mean square error and the residual degrees of freedom. Section 5 covers the effect of covariate balance through the variance inflation factor. This is the key to understanding what the value if any is of stratification. Section 6 presents a brief example. Section 7 is a discussion of the results and section 8 provides some recommendations,

# 2 A three aspects framework

When considering the standard linear model, there are three effects on precision of a treatment effect estimate due to fitting a covariate[7]. First, there is the *mean square error (MSE) effect*. To the extent that the covariate is prognostic, the expected value of the MSE to be used in calculating variances of estimates of treatment effects will be reduced. Second, there is *the imbalance effect*. To the extent that the covariate is imbalanced between the treatment arms, the variance will be increased and this is allowed for in regression approaches by multiplying the estimated variance by the appropriate *variance inflation factor*(VIF)[8] or non-orthogonality penalty [9, 10]. Third there is the *second order precision effect*. A degree of freedom will be lost in estimating the MSE.

The relationship of these three factors to precision is very different and this makes it desirable when considering them to separate their different roles. When the decision is made whether or not to use the baseline in an ANCOVA, the prime motivation, of course, is the expected reduction in the estimated MSE by fitting the baseline. This effect is usually much more important than the other two. However, as we discuss below, this factor is *not* relevant when making a decision to stratify or not. For this, it is the VIF that matters. In consequence, since the question that is addressed in this work is the value or otherwise of stratification and what analysis to use if one has stratified, it is the VIF on which we shall concentrate attention.

It is important to understand the relationship the three factors have to the model, the design and the outcome. These are summarised in Table 1.

| | | Effect | | |
|---|---|---|---|---|
| | | Mean square error effect | Imbalance effect or VIF | Second order Precision |
| Influence | Design | No | Yes | No |
| | Model | Yes | Yes | Yes |
| | Outcome | Yes | No | No |

*Table 1 Influence of the three factors, Design (stratified randomisation or simple randomisation), Model (linear models without or with adjustment for different covariates) and Outcome (different correlations of the outcome variables and the covariates) on the three effects on precision.*

Note that although all three effects are influenced by the model, as regards second order precision, it is only the *rank* of the model that matters, since this determines the number of degrees of freedom lost for estimating the error term.

As regards the variance inflation factor, it is not affected by the outcome, because we restrict our attention here to the simple linear model with estimation using ordinary least squares. Under such circumstances, as is well known in the literature on design of experiments, design efficiency only depends on the design matrix[11]. For extensions of the linear model, this does not apply and the efficiency of chosen designs depends on unknown parameters that have to be estimated[12], perhaps by iterative reweighting. For example, for binary outcomes the variance of a response depends on its expected value and for hierarchical models, such as might be applied to incomplete blocks designs, on the variance covariance matrix. These parameters have to be estimated using the outcomes.

Finally, what the table states about the mean square error effect is true about the *estimated* mean square error effect given an appropriate model . An *appropriate model*, in this context, is a model that includes every prognostic factor whose distribution has been restricted (as opposed to being randomised.) Further unrestricted covariates that are prognostic may be included and restricted covariates may be excluded if not prognostic given other factors in the model. If this is the case, only the model is relevant for considering the effect on MSE. However, if one restricts allocation in some way that is *not* reflected in the ANCOVA model, the actual mean square error is reduced (if the covariate is prognostic for the outcome) but the estimated mean square error from the ANCOVA model will be biased and larger than it should be, a point which RA Fisher raised some 90 years ago in his discussion[13] of Student's non-randomised designs[14, 15]. Indeed , in a recent paper using simulations to examine strategies for analysing stratified designs Sullivan et al state[16], 'Since stratified randomisation induces a correlation between treatment groups, failure to adjust for stratification variables when estimating treatment effects can lead to overly wide confidence intervals, type I error rates less than $\alpha$, and reduced statistical power.'(p2) As discussed above, if one stratifies, one should adjust in the analysis. However adjustment is possible whether or not one has stratified (provided that the covariate has been measured) and it also follows that if one has decided to adjust, stratification of itself plays no further role as regard two of the three effects summarised in Table 1 above: the *MSE effect* and the *second order precision effect* are the same. For example, if, not having stratified, one simply recoded a covariate as 0 or 1 depending on whether its value was below or above the median, as Sullivan et al did in their simulation (section 3.1) and adjusted for it, the expected consequences for these two effects would be the same as if one had stratified. This then raises the question, *what is the value of stratification*? As we shall show, the key to understanding the value of stratification and the consequence of fitting various models is the variance inflation factor. We shall discuss this in due course. However, first we divert to discuss briefly the effect of the other two factors.

# 3 The expected mean square error effect

This has been often discussed. (See, for example p16 of the famous paper by Hills and Armitage[17].) We can consider this by supposing that the continuous predictor $X$ has been standardised as has the outcome variable $Y$. The regression of $Y$ on $X$ is then the same as the correlation between the two, $\rho$, and, asymptotically, the reduction in the variance of the estimate of the treatment effect by fitting Model (B) rather than Model (A) is given by the ratio $1-\rho^2$. A very common covariate to consider is the baseline value corresponding to the outcome and it is then interesting to compare covariate adjustment to simple analysis of the *change-score* (sometimes called the *gain-score*), that is to say, the difference of outcome from baseline. This is illustrated in Figure 1 which is based on Figure 7.4 of *Statistical Issues in Drug Development*[9] and takes as a reference analysing the 'raw' outcomes alone and ignoring the baseline altogether.

Note that, whatever the correlation coefficient, the covariate adjusted analysis always yields a lower or equal variance of the treatment effect estimate. However, the result is asymptotic. The fact that an extra parameter, the correlation coefficient, has to be estimated means that there is some loss to which neither alternative analysis (raw outcome or change score) is subjected. This suggests that it could be possible that for very low value of the correlation coefficient, raw outcomes could be preferable. Similarly, for high values, the change score could be preferred. Understanding this point requires looking at the other two factors in Table 1, second order precision and the variance inflation factor.

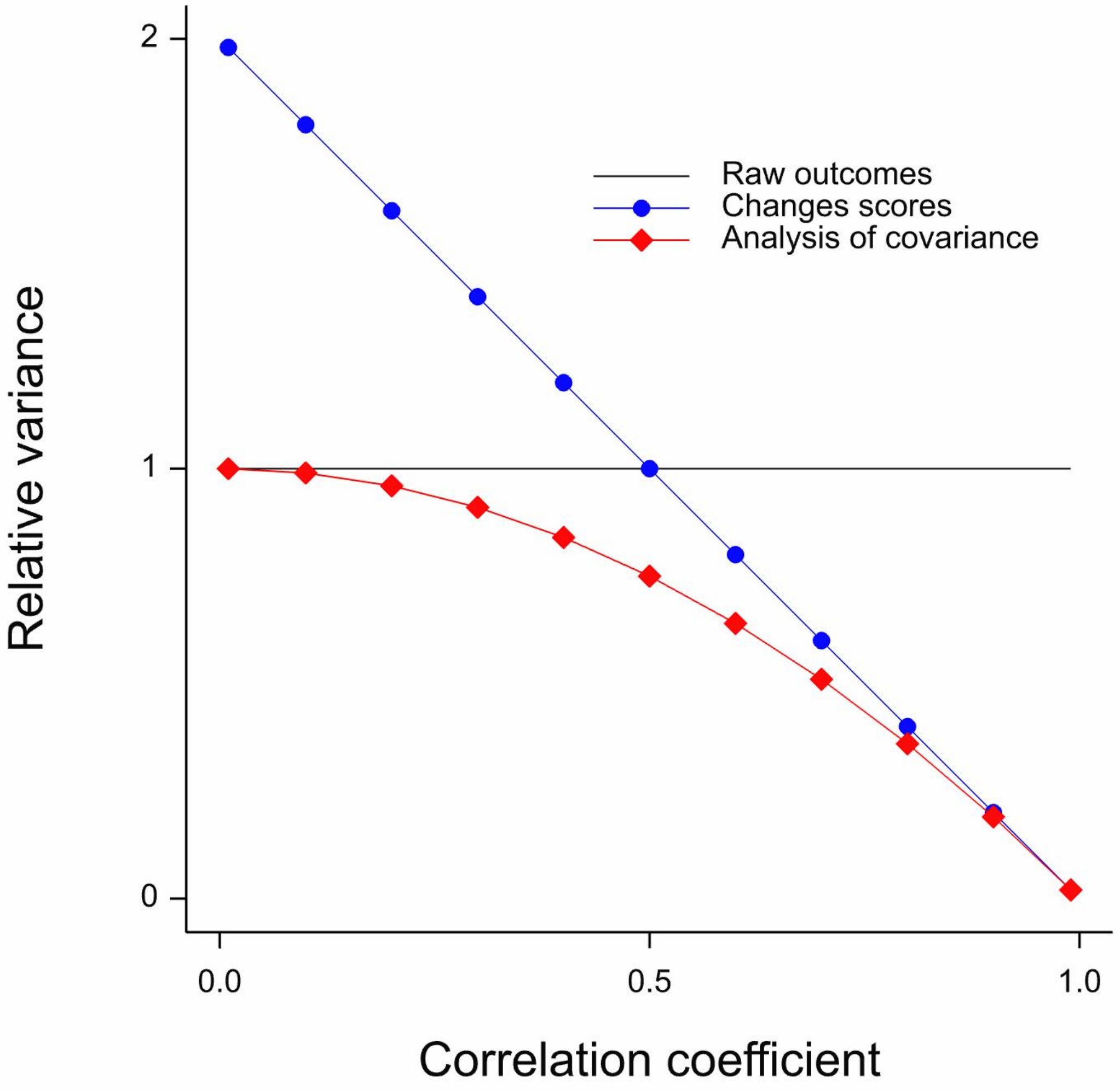

*Figure 1 Variance of the treatment effect estimate for Model(B) (ANCOVA using X) as a function of the correlation coefficient compared to Model (A) (analysis of raw outcome only) as well as the common strategy of using a change score.*

# 4 Second order precision

When an extra term is fitted, the residual degrees of freedom for estimating the residual variance are reduced by one to the extent that the fitted term is not redundant. That is to say, if there are $N$ patients in total and a matrix of predictors, $\mathbf{P}$, the degrees of freedom $\nu$ are given by $\nu = N - rank(\mathbf{P})$. We thus have $\nu = N - 2$ for Model (A), where a degree of freedom is lost for the treatment indicator in addition to the intercept, $\nu = N - 3$ for Models (B) and (C), where a further degree of freedom is lost either for $X$ or for $S$ and $\nu = N - 4$ for Model (D), where two degrees of freedom are lost for fitting both.

For any type-one error or level of confidence chosen, the residual degrees of freedom will determine the critical values of the t-distribution that has to be used. However, a way of indexing second order precision that is not linked to any given level chosen, is to use the variance of the t-distribution, which is

$$Var(t_\nu) = \frac{\nu}{\nu - 2}, \nu \geq 3.$$

Given this formula all that is necessary is to substitute the relevant values for $\nu$, that is to say $N-2,3 \text{ or } 4$ as the case may be. The results as functions of $N$ are shown in Figure 2,

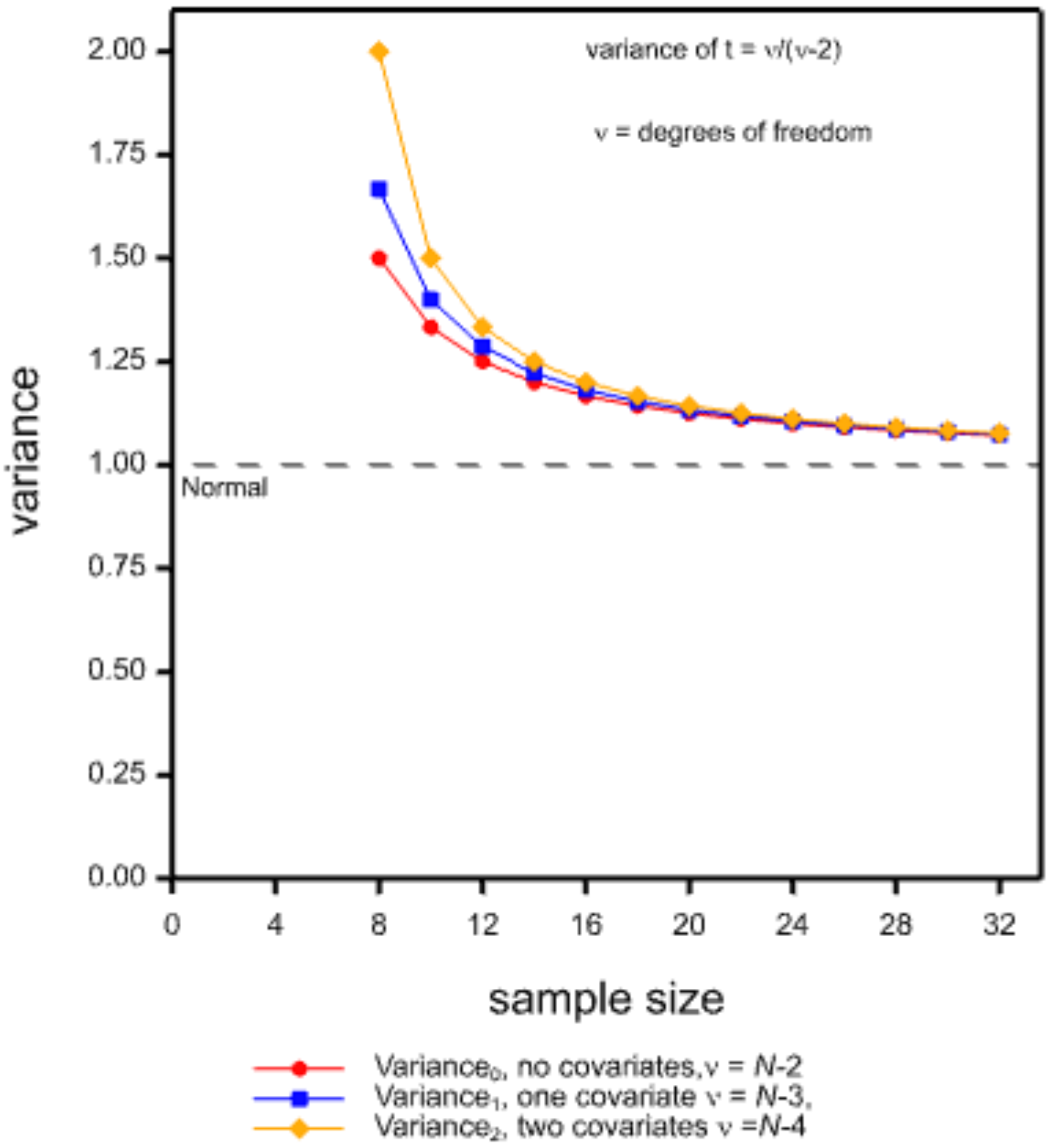

*Figure 2 Variance of the t-distribution for various Models A)-D).*

Since it may be of interest to compare the variances of the different models an alternative representation is in term of the ratio of the variances compared to the model with no covariates. This is given in Figure 3.

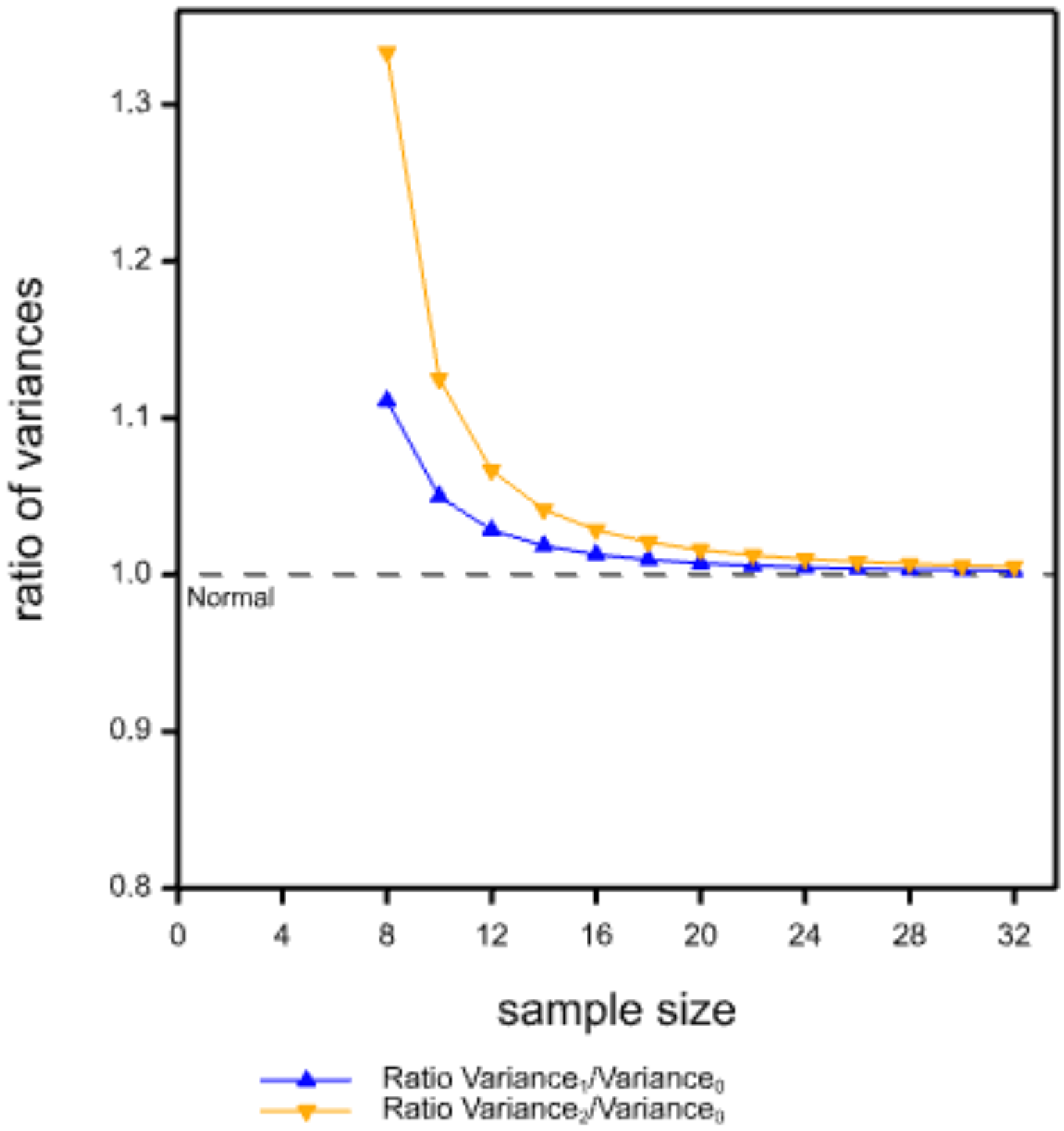

*Figure 3 Ratio of variances of the t-distribution fitting one (Variance$_1$) and two covariates (Variance$_2$) compared to fitting none (Variance$_0$) for various sample sizes.*

# 5 The variance inflation factor and the value of stratification

We regard the VIF as central to understanding why exactly it is that balance matters. Various recent reviews of covariate adjustment stress that balance is valuable without really explaining why. For instance, Van Lancker et al[18] in a thorough investigation of many consequences of using ANCOVA state, 'In finite samples, adjusting for covariates with a very weak or null association with the outcome may not lead to efficiency gains and can even result in a precision loss' and with respect to stratified biases coined designs, 'Such designs create balance of treatment arms with respect to the stratified variables in order to improve efficiency of the treatment effect estimate' . However, they do not pinpoint or quantify why and how balance contributes to efficiency. Here we explain how the VIF holds the key to understanding this.

If it has been decided to fit Model (B), the value of stratification is that it reduces the non-orthogonality penalty due to imbalance. Qu provides a careful examination of the effect of stratification on covariate balance[4]. However, what we are suggesting here is to take it one step further. Rather than considering the effect of balancing strategies on the covariate itself, we should consider how balance or imbalance effects the variance of the treatment effect estimate. If the covariate is in the model, its effect is adjusted for: the treatment effect estimate is unbiased and the expected mean square error is reduced, whether or not the factor is unbalanced, to the degree that it is prognostic. Furthermore, the residual degrees of freedom are unaffected by the degree of imbalance. What changes is the variance inflation factor. This is what we examine in this section,

For a trial with *n* subjects per arm $i, i = 1, 2$ , and thus $N = 2n$ in total, in which there is adjustment for a linear predictor, *X*, the *variance inflation factor* (VIF) will be:

$$VIF = \frac{\sum_{i=1}^{2}\sum_{j=1}^{n}\left(X_{ij} - \bar{X}_{..}\right)^2}{\sum_{i=1}^{2}\sum_{j=1}^{n}\left(X_{ij} - \bar{X}_{..}\right)^2 - \frac{n}{2}\left(\bar{X}_{2.} - \bar{X}_{1.}\right)^2} \tag{1}$$

(See appendix.) This is the factor by which the usual multiplier for the MSE of $2/n$ must be further multiplied to produce the variance of the treatment effect. Unless $\bar{X}_{1.} = \bar{X}_{2.}$ , which is to say unless the mean values of the covariate are identical in the two arms of the trial, this factor will be greater than one. The value of stratification is that it forces the expected value of this difference to be smaller than it otherwise would be.

If the covariate is Normally distributed, then in the absence of stratification, the expected value of the imbalance effect can be shown to be[19, 20]

$$E\left[VIF\right] = 1 + \frac{1}{N-4} = \frac{N-3}{N-4}. \tag{2}$$

(This formula applies if the covariate is Normally distributed and was given by Cochran[19] in 1957. It is derived in an appendix because it will be modified below for the case where the covariate has been stratified.) For moderately large values of $N$ this is quite modest, a fact that is possibly not well understood. For example, for $N = 200$ , which would be a small size for a Phase III clinical trial, one has $E\left[VIF\right] = 1.005$ to three decimals places, so that the expected variance inflation would be about ½ of one percent. This calls into question the value of stratification and also other balancing strategies such as minimisation for such trials.

It is hard to give general rules for what benefit stratification will bring but one will have

$$1 \le E\left[VIF_S\right] \le \frac{N-3}{N-4} = 1 + \frac{1}{N-4},$$

where $VIF_S$ is the variance inflation factor value given stratification.

Sullivan et al[16] simulate covariate values, $X$, from a standard Normal distribution, stratifying at $X=0$. (See section 3.1 of their paper.) The effect of this can be judged approximately by considering that the variance of a Normal distribution truncated at 0 is $1-2/\pi$, which is 0.363 to three decimal places (see appendix). In each stratum, values can be regarded as being sampled from a distribution with this variance and this means that whereas such means have a variance proportional to 1 if sampled from the parent Normal distribution, they have a variance proportional to 0.363 when sampled from the truncated distribution. The resulting distribution is, however, not itself Normally distributed, so that standard distribution theory does not apply. Nevertheless, a value of

$$E\left[VIF_s\right]=1+\frac{1-2/\pi}{N-4}\approx 1+\frac{0.363}{N-4} \tag{3}$$

should apply approximately, although it must be understood, that in practice it will not be possible to predict perfectly when recruiting patients what the median is for the population from which theoretically they could be considered to be drawn.

It is also interesting to consider the effect of stratification on using Model (D). The following argument can be used. If, having stratified, we fit Model (C), there will be no loss of orthogonality since we are only fitting the stratum indicator, which is balanced by design. If we now add as a further covariate the continuous predictor $X$, then it is the component of $X$ that is orthogonal to $S$ that is relevant to determining the VIF. However, given that it is orthogonal the expected penalty is that indicated by (2) but adjusted for the fact that we have lost one further degree of freedom for fitting the stratum. The formula for the penalty is

$$E\left[VIF_{S,X}\right]=1+\frac{1}{N-5}. \tag{4}$$

One further formula may be of interest and that is the inflation factor that applies to a completely *randomised* design fitting Model (D), that is to say not only the continuous covariate $X$ but also a dichotomy of it at the median even though stratification has not taken place. The problem is that there is a finite probability that the allocation will be completely confounded. For a trial with $2n$ patients, and subject to the constraint that there are $n$ patients per arm, there are

$$\frac{(2n)!}{n!n!}$$

possible allocations. Of these allocations, there are two for which either all values for group 1 (say) will be above the median and all those for group 2 will be below the median or vice versa. Thus, the probability of complete confounding is

$$2\frac{n!n!}{(2n)!}. \tag{5}$$

For moderately large values this becomes very small. For example, for $n=10$, that is to say 20 patients in total, it is less than one in 92,000. Nevertheless, a purist might claim that this means that the expected values of the VIF is undefined, since there is a finite probability of an infinite variance. Note that, in theory, a similar problem could arise with median stratification, since the 'median' used for allocation could (unbeknown to the trialist when the trial was designed) be higher (or lower) than the value for any patient. Then, effectively, there would only be one stratum and the baseline values would be as if completely randomised. Obviously, however, this is unlikely. In any case, this is not a problem for validity of marginal modelling inferences, since the stratum indicator can be eliminated from the model and since that choice will have been made based on covariate distribution only, inferences will not be biased. It will, of course, be a problem for conditional inferences.

If this objection can be ignored, then for moderately large samples we can treat both covariates as if they were Normally distributed. A further paper of ours treats adjustment for any number of covariates, both binary and continuous and shows that the general formula given by Cox and McCullagh[20] in 1982 (p547) can be applied . For $k$ covariates we have that

$$E\left[VIF(k)\right]=1+\frac{k}{N-k-3}. \tag{6}$$

may be used as an approximation, even though one of the predictors is binary. Substituting $k=2$ we have

$$E\left[VIF(2)\right]=1+\frac{2}{N-5}. \tag{7}$$

Expected variance inflation factors for models B and D for both stratified and randomised designs are given in Figure 4 Figure 4 Variance inflation factors for randomised (red) and stratified (blue) designs, theoretical values (lines) and empirical (symbols) results for models B and D (10^6 simulation runs per scenario).as are means from $10^6$ simulation runs sampling the covariate from a Normal distribution. The theory seems to hold up well. Of course, one can argue that the Normal distribution might not accurately cover any given covariate distribution. However, there are three answers to this. The first is that the Normal distribution is what others have assumed in simulating. If such simulations are justified, it is surely of interest to look at what theory has to say. The second is that considering the Normal distribution is a start. It is useful to have established what the theory says about this standard case. The third is that the stratum indicator itself, which is included in model D, is not Normally distributed but the theory appears to hold. In fact, our investigations of this point, which will be presented in a further paper on this topic, suggest that for moderate sample sizes, even for binary covariates the theory holds up quite well.

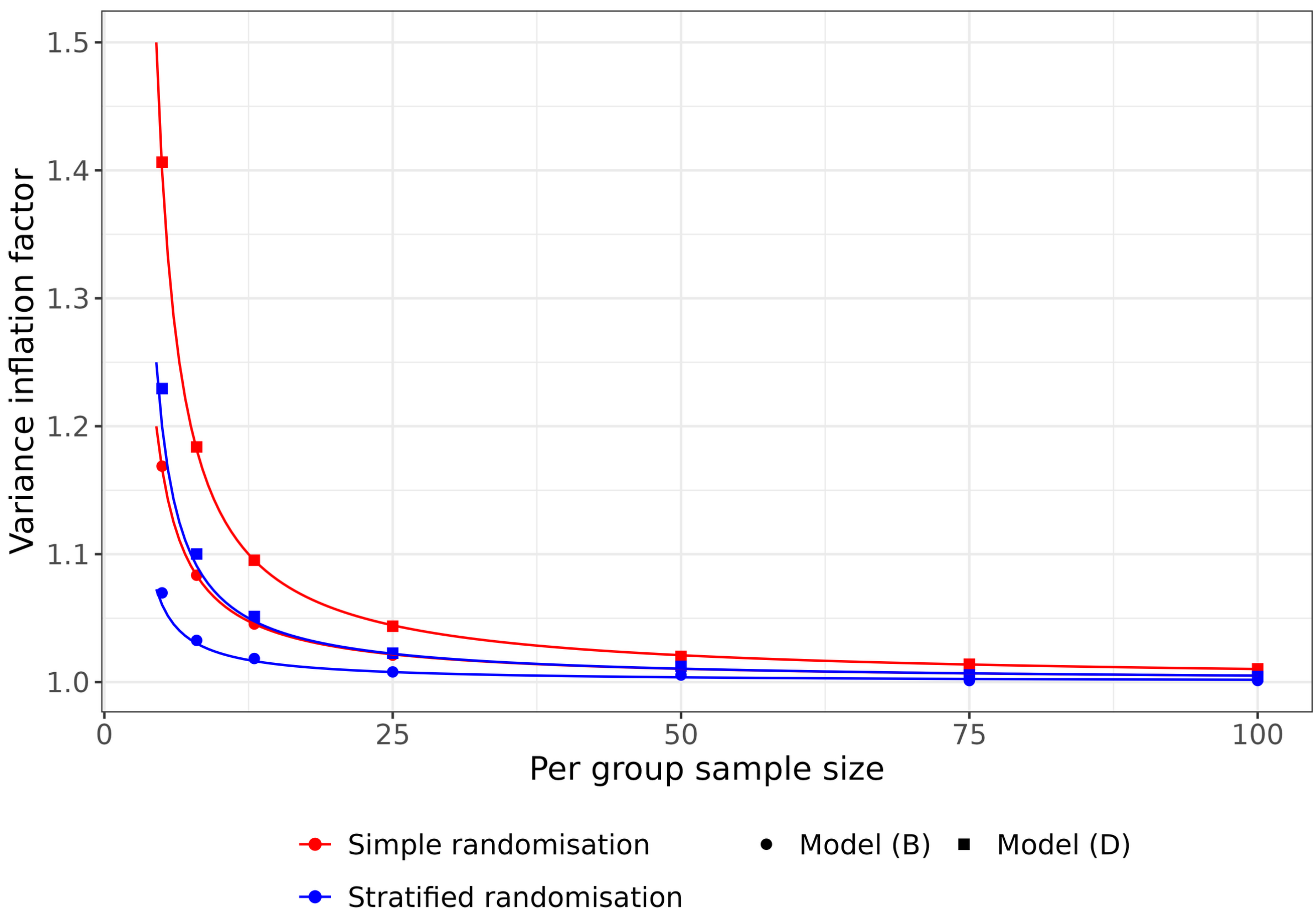

*Figure 4 Variance inflation factors for randomised (red) and stratified (blue) designs, theoretical values (lines) and empirical (symbols) results for models B and D (10^6 simulation runs per scenario).*

# 6 An Example

We present a small example that has been adapted from data given in Table 12.9 of Piantadosi's famous book[21] and which concerns a trial in familial adenomatous polyposis (FAP) that compared sulindac to placebo[22]. We have reduced the data to that for 16 patients only in order to provide an example that permits division of patients into four equally sized groups produced by median stratification and treatment. The trial may seem to be rather small but Hee at al, in their survey of trials in rare diseases[23], found a median size of 15 for phase II trials. One value has been slightly modified to avoid ties compared to the median. The data we shall use are polyp size at baseline. Since we are interested in the VIF, the values at outcome are not relevant.

The data are given in Table 2. We resampled from these data in two ways. First by considering all possible allocations that would give eight patients on each treatment. Second, by creating two strata by noting whether values are above or below the median and then allocating four patients to each treatment in each stratum. The former approach yields $16!/(8!8!) = 12870$ possible allocations. The latter approach yields $\left(8!/(4!4!)\right)^2 = 70^2 = 4{,}900$ possible allocations. We have simulated 5,000 samples from the former and performed a complete enumeration of all 4,900 possible allocations for the latter.

| Patient ID | Treatment | Baseline Size |
|---|---|---|
| 12 | Sulindac | 1.7 |
| 8 | Placebo | 2.0 |
| 7 | Sulindac | 2.2 |
| 14 | Placebo | 2.3 |
| 15 | Placebo | 2.4 |
| 13 | Placebo | 2.5 |
| 5 | Sulindac | 3.0 |
| 16 | Sulindac | 3.0 |
| 22 | Sulindac | 3.1 |
| 4 | Placebo | 3.4 |
| 23 | Sulindac | 4.0 |
| 6 | Placebo | 4.2 |
| 9 | Placebo | 4.2 |
| 10 | Placebo | 4.8 |
| 3 | Sulindac | 5.0 |
| 11 | Sulindac | 5.5 |

*Table 2 Modified data from a trial in FAP reported by Piantadosi showing polyp size at baseline for 16 subjects. The data have been sorted by polyp size. The value for patient 22 has been increased from 3.0 to 3.1 in order to yield a unique median.*

Table 3 below gives the results of the calculation of the VIF for treatment for resampling for this example.

| Allocation method | Number of allocations | Expected | Observed Mean | Observed Median | Standard Error of Mean |
|---|---|---|---|---|---|
| Completely Randomised | 5000 | 1.083 | 1.086 | 1.035 | 0.0021 |
| Stratified | 4900 | 1.030 | 1.022 | 1.011 | NA |

*Table 3 Results for the VIF obtained by taking the baseline values as fixed and re-allocating treatment. The results for completely randomised allocations reflect a true simulation. For the stratified case, the VIF has been calculated for every one of the 4,900 possible allocations. The result is thus not subject to sampling error and the standard error of the mean is not quoted.*

It can be seen that the theoretical expected values, which strictly speaking apply to sampling from a Normal distribution, agree fairly well with the empirical results based on resampling. It can also be seen that there is a modest reduction in the VIF as a result of median stratification. Figure 5 gives density plots for the two types of allocation. It can be seen that the distributions are highly skewed towards large values, which explains that the mean VIF is larger than the median for both types of allocation. Complete randomisation is much more highly skewed than median stratification.

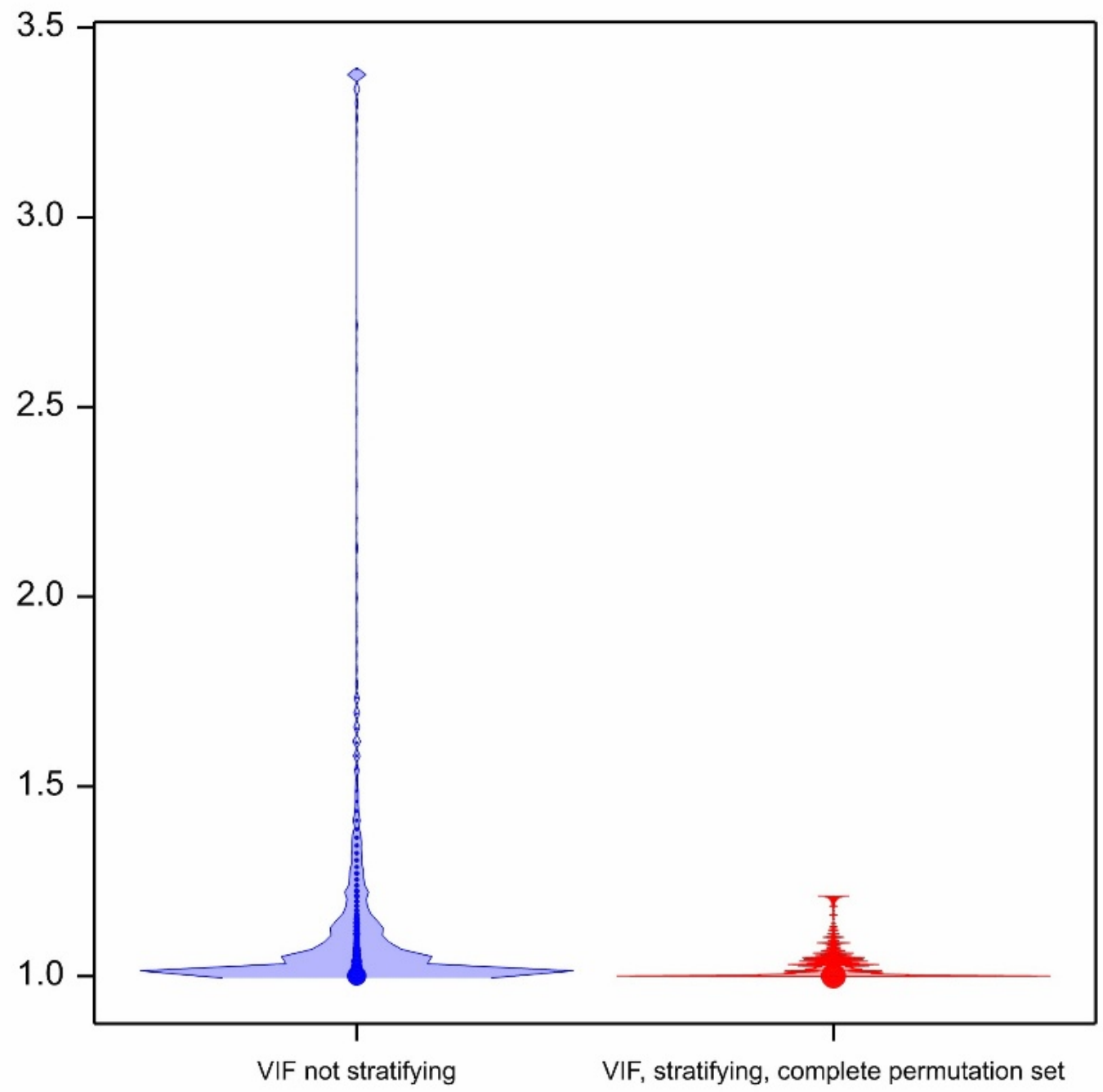

*Figure 5 Density plots for VIFs for randomisation (not stratifying) and stratification. The randomisation values are based on a simulation with 5,000 re-samplings. The stratification distribution is based on complete evaluation of all 4,900 cases.*

# 7 Discussion

As was pointed out in *Statistics in Medicine* 30 years ago[24] (p1721), 'having decided to condition, we minimize the standard errors of our treatment effects if our covariate is orthogonal to treatment. This is the true value of balance, which is an issue of efficiency, not validity.' This ought to be more widely understood and it also ought to be understood that the potential for increasing efficiency in this way is limited. What is important is to condition on that which is prognostic.

It is interesting to note that an early paper on stratification for a continuous covariate by Finney considered trying to make *mean* covariate values close[25]. This is the relevant target for a continuous covariate for minimising the VIF. This was possible in the animal studies he was considering, where the animals to be treated were available as a group before starting the study. However, for clinical trials this will rarely be possible and the sort of stratification considered here will have to be used. As Sullivan et al[16] point out, given that patient accrual has to be sequential in clinical trials, balance of total numbers and, by extension, balance in any stratum, can only be achieved by some device such as permuted blocks[3, 26].

Thus, usually, subjects will be balanced by blocks that are smaller than the strata. Yet, typically one does not adjust for this extra balancing factor: 'block' will not be in the model[26]. A justification for not putting block in the model is that it is just a device for balancing the dichotomised covariate represented by the strata and not expected to have any prognostic value of its own. The exception to this would be if it did have a prognostic value because some time trend was expected[26, 27].

Thus, if block can be ignored when balancing strata, because it is not expected to bring additional prognostic value, it would seem logical that stratum can be ignored if it is not expected to bring additional modelling value to the covariate from which it was created. In other words, it would be logical to use Model (B) of Qu's[4], rather than Model (C). Another argument can be given as follows. Although there is no penalty in terms of VIF for fitting Model (C), as expression (3) shows, the penalty for fitting Model (B) is small. It is likely, however, that as regards expected MSE, whatever the true relationship is between the baseline and the outcome, it is likely to be better approximated by a straight line than a step-function. This agrees with Sullivan et al[16] who state, 'We therefore advocate adjustment for the underlying continuous values of stratification variables rather than adjustment for the categories used for stratification.' (p10). As regards Model (D), things are not quite so clear and it is useful to consider what the consequences will be of fitting this model, rather than Model(B) on the three factors we have identified as relevant. For large trials, the advantage might lie with D but for small trials with B.

The second factor, the VIF, will be increased in expectation from expression (3) to expression (4). The third, second order precision will be affected by the loss of one degree of freedom. The first, mean square error might be reduced if the regression of $Y$ on $X$ is not perfectly linear. For example, one might suppose an unidentified form that might be approximated by some modelling scheme such as splines[28], fractional polynomials[29, 30], orthogonal polynomials[31] or regular polynomials. As an example of the latter, it is interesting to consider the correlation of the stratum indicator with various powers of $X$ . In order to make progress in developing some simple theory we shall make the following assumption that can apply only approximately at best, namely that the sample is very well behaved as regards the stratification plan and that all standardised values below the median are negative and all those above the medians are positive. Where this is the case, we have.

$$S = \begin{cases} -1 & X < \tilde{X} \\ 1 & X > \tilde{X} \end{cases}, \tag{8}$$

Where $\tilde{X} = 0$ is the median of $X$ The values are given in Figure 6 and have been calculated as follows. We note that a positive and negative stratum indicator will be associated with equal probability with negative and positive values of $X$ . For any even power of $X$, the expectation is thus zero, since an even power will be positive and associated with a negative or positive stratum indicator with equal probability. However, a negative value of an odd power will always be associated with a negative stratum indicator and a positive one with a positive stratum indicator, so the expected product of indicator and power is given by the corresponding moment of the standard Normal distribution truncated at zero. As is explained in the appendix, this may be adapted to provide expected correlations between the stratum indicator and the odd powers of $X$ .

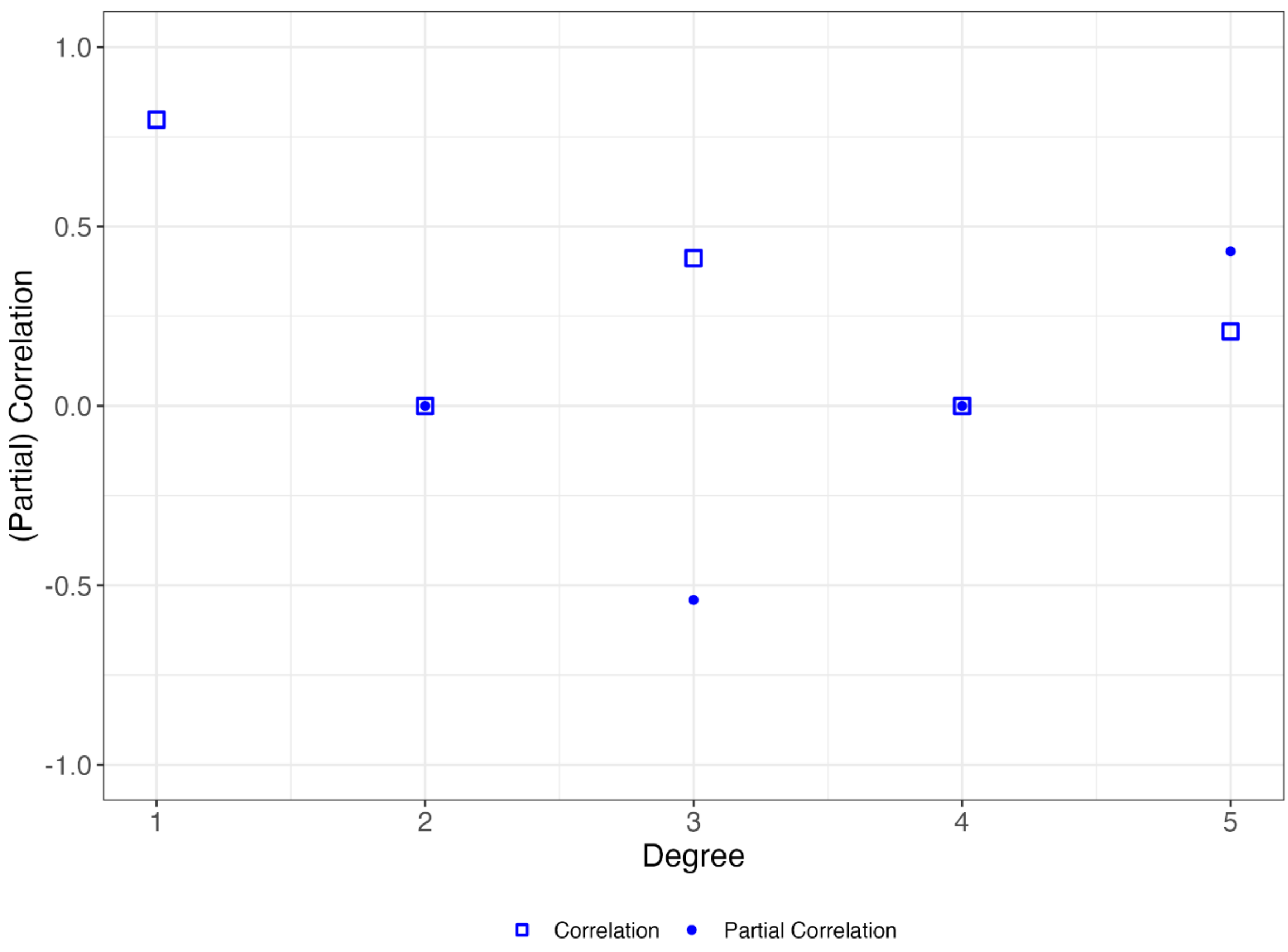

*Figure 6 Correlation of the stratum indicator with various powers of X (squares) and the partial correlation (dots) of the stratum indicator with various powers of X given X. For even degrees, both the simple and partial correlations are 0 and the dots overlap.*

If Figure 6 is examined, it will be noted that the correlation with all even powers of $X$ is zero and that for each successive odd power of $X$, the correlation is about one half of the previous odd power. However, as we saw when moving from Model (B) to Model (D), it is really the partial correlation with the treatment indicator that is relevant. Many schemes for models are possible and we have already mentioned some. However, in a randomised design, a scheme in which successive polynomial terms to order $m$ were used would require a value of $k = m$ to be used in expression (6). It seems intuitively reasonable that however much could be achieved in reducing the VIF by median stratification, the maximum effect would equivalent to setting $k = m - 1$ and that this is unlikely to be achieved in practice.

We prefer not to be proscriptive regarding the choice between Model(B) and Model(D). We suspect that in many cases choosing D rather than B will bring little by way of benefit but also will not cause much harm. Whatever choice is made, we advise that it should be agreed and prespecified.

So far we have not dealt with situations in which the treatment-by-baseline interaction is considered important. Regulatory guidance has generally stressed using a model without an interaction for a primary investigation and looking at interactions in a secondary one[32, 33]. Models that do not include interactions are generally robust even if an interaction is present but since they pool the interaction term with the error term in the residual mean square, may lose some power. That is a matter, however, that is not directly related to this investigation. However, in the discussion we provide a formulae that may be used for models with interactions.

As regards the value of median stratification as opposed to randomisation if, say, model(B) is envisaged, our opinion is that this might be valuable for a small trial, such as the one illustrated in section 6. Comparison of the VIFs in Figure 5 for the small trial we considered support this. (Note however, that stratifying on the true median as opposed to the predicted median, which would have to be used in practice, overstates the value.)  For large phase III trials it will be of minimal advantage. For example, for 50 patients the expected value of the VIF given randomisation is 1.022 and for a stratified design is 1.008.

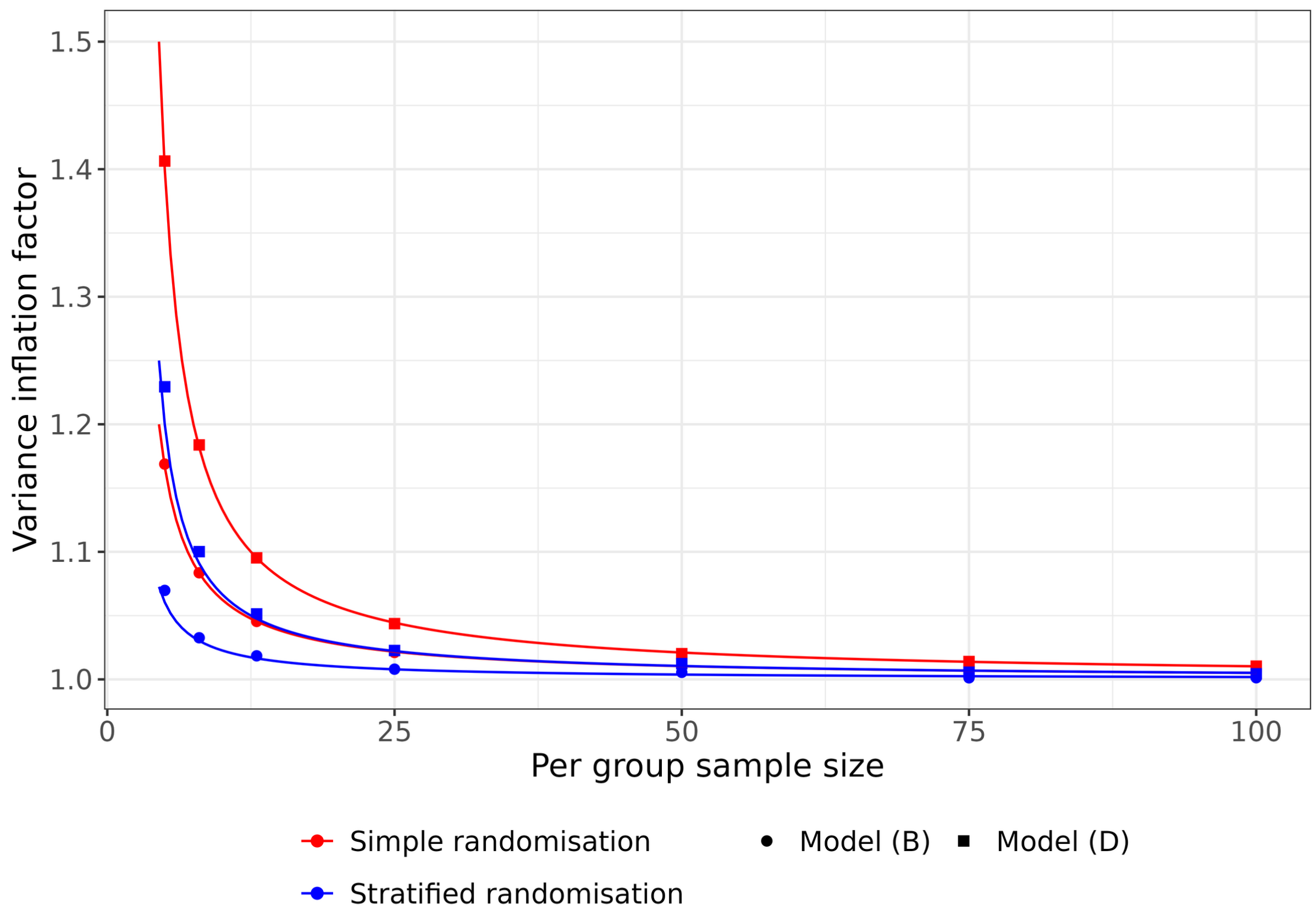

may help trialists decide whether stratification is worthwhile.

The theory we have given applies to stratification for one covariate. Given that the context we have considered is rare diseases, fitting more than one covariate may be rather ambitious. Formula (6) suggests the VIF that will apply.  However, one possible strategy is to construct a prognostic score from historical data[2, 7, 34]. Allocation could then be stratified by this score, in which case, the theory we have presented applies. A further paper of ours considers expected values and variances of VIFs given randomisation when many covariates are fitted but we do not think this is of much practical relevance to the rare disease situation.

Trials with three or more arms raise other interesting considerations. Variance inflation increases as the number of columns in the design matrix increases relative to the number of rows. Now consider a trial with $t \geq 3$ arms, a single covariate and $n$ patients per arm, where $n$ is small, If, when comparing two treatments. the data from other treatments are ignored, the design matrix is reduced from having $1+\left(t-1\right)+1=t+1$ columns, where the three terms on the left hand side are respectively the degrees of freedom for intercept, treatments and covariate to having three columns only, However the rows of the matrix are reduced from $tn$ to $2n$ and the net effect will be to increase the VIF. It may thus be advantageous to use the data from all treatments, imposing the assumption of common slopes across all treatments, including ones not involved in a given contrast.

However, as the value of $n$ increases, such an assumption of homogeneity of slopes may do more harm than good, and Ye at al have suggested that allowing for heterogeneity may be beneficial where sample sizes are large and trials have three or more arms[35]. A similar issue arises with variance estimation. In analysing clinical trials, many use a linear model that pools the variance estimates from all treatments. As was pointed out by Senn[26], this makes sense in the context of agricultural field trials, where degrees of freedom are scarce but less so for large phase III clinical trials. We consider, however, that the large multi-armed trials are not really relevant to the rare disease context we have considered.

As we explained in the introduction, we have limited our investigation to main effect models. However, the three factor framework can be extended to models with treatment by covariate interactions. For example we might be interested in models

$$\begin{aligned} &Model(Bi): Y \sim Z + X + Z \times X \\ &Model(Ci): Y \sim Z + S + Z \times S \quad . \end{aligned}$$

where the previous Models (B) and (C) have been modified to include interaction terms indicated by $\times$ . To the extent that our stratification strategy has succeeded in balancing perfectly by the stratum indicator, model (Ci) incurs no variance inflation and the effect of fitting the interaction is simply to remove the interaction from the residual mean square error at the cost of losing one degree of freedom for second order precision. Equivalent mean square error and second order precision effects will also occur for Model (Bi). However, for this model there will be an increase in the variance inflation factor compared to Model (B). We can argue as follows. To the VIF for model (B) one degree of freedom has to be removed from the denominator because an extra term is being fitted. However, what is relevant to the VIF numerator is that a further element orthogonal to the covariate already fitted must be accounted for. We thus need to modify the term $\left(1-2/\pi\right)/\left(N-4\right)$in (3) to become $\left(2-2/\pi\right)/\left(N-5\right)$and thus obtain,

$$E\left[VIF_{Bi,S}\right] = 1 + \frac{2-2/\pi}{N-5} \approx 1 + \frac{1.363}{N-5} \tag{9}$$

where $VIF_{Bi,S}$ is the VIF for $Model\left(Bi\right)$ given stratification. Our simulations suggest that it works well. Our feeling is that given the small sample sizes such models will not usually be of interest. If stratification has not been applied but a completely randomised design has been used, then the relevant formula for the VIF is

$$E\left[VIF_{Bi,R}\right] = 1 + \frac{2}{N-5}. \tag{10}$$

This is the same formula as expression (7) However, others who disagree may find consideration of expression (9) useful.

the reader who is interested in advice on models with interactions allowing for heterogeneity should consult the papers by Ye et al, Wang et al and the paper by Van Lancker et al already cited[18, 35, 36]. We are assuming that the average treatment effect is of primary interest even if an interaction is fitted. Looking for subgroup effects has the consequence of making a rare disease even rarer.

# 8 Recommendations

Finally, we suggest the following way of looking at covariate adjustment and design of clinical trials.

1) The intended analysis guides the design. The main reason for balancing a covariate is to reduce the non-orthogonality penalty, which is to say, to make the VIF closer to 1 than it otherwise would be, given that the covariate is in the model.
2) Since the intended analysis guides the design and since the reason for stratification is to make this intended analysis more efficient (by balancing to the extent possible), one is not committed to putting the strata in the model.
3) However, one is committed to putting the covariate in the model, since that is why the design was stratified. Thus Model(B), which is the approach recommended by Sullivan et al[16] in their discussion is a consistent approach.
4) Nevertheless, stratifying can contribute to reducing the conditional bias given an observed distribution of a covariate, $X$ if the effect of the covariate is not approximately linear. This is a point that Qu has made[4] (P235). Under such circumstances it will also have an effect on the true standard error of the estimate. To be consistent. such a reduction ought to be reflected in the estimated standard error for the treatment effect and this can be achieved by also fitting the stratum indicator. Thus, using Model (D) can also constitute a consistent position.
5) However, one is neither limited to Model(B) nor to Model(D) and one should use a form (transformation etc) of the covariate that it is believed is valuable prognostically.
6) Given that the intended analysis guides the design and given that pre-specification increases confidence in the results obtained, the intended form of adjustment should be pre-specified, as is required in any case by regulatory guidance[37].
7) We have suggested that in the context of rare diseases, fitting the covariate interaction will usually not be of interest. However, if this is done our advice is that users should take care that the definition of the treatment effect is that for the patient with the average baseline score and that the model is fitted accordingly. See the discussion in the Many Modes of Meta[38] pp542-545.

# Acknowledgement

We thank our colleague Robin Ristl for drawing our attention to the paper by Qu[4] and Tim Morris for helpful comments on earlier versions of our paper. FK and MP are members of RealiseD, which is supported by the Innovative Health Initiative Joint Undertaking (IHI JU) under grant agreement No 101165912. The JU receives support from the European Union's Horizon Europe research and innovation programme and COCIR, EFPIA, Europa Bío, MedTech Europe, and Vaccines Europe. Views and opinions expressed are those of the authors only. This publication reflects the author's views. Neither IHI nor the European Union, EFPIA, or any Associated Partners are responsible for any use that may be made of the information contained herein.

# Appendix

## Variance Inflation Factor (VIF)

We suppose for simplicity that we have *n* patients per arm with $N = 2n$ in total. The covariate values may be indicated by $X_{ij}, i = 1,2, j = 1 \ldots n$, where $i = 1,2$ indicates treatment group membership. The treatment dummy variable may be coded as $-1/2, i = 1, 1/2, i = 2.$ Without loss of generality the covariate may be recoded in terms of deviations from the overall mean as $x_{ij} = X_{ij} - \bar{X}_{..}.$

The treatment indicator and recoded covariate sum to zero and are therefore orthogonal to the intercept indicator, which we may therefore ignore. We can now formulate the predictors as a $2n \times 2$ matrix, where the first column carries the treatment dummy variable and the second the covariate. The matrix of sums of square and cross-products is

$$\mathbf{x}'\mathbf{x} = \begin{pmatrix} \frac{n}{2} & \frac{n}{2}\left(\bar{x}_{2.} - \bar{x}_{1.}\right) \\ \frac{n}{2}\left(\bar{x}_{2.} - \bar{x}_{1.}\right) & \sum_{i=1}^{2}\sum_{j=1}^{n} x_{ij}^2 \end{pmatrix}.$$

If this is inverted, then the element of the first row and column will be found to be $VIF \times \left(2/n\right)$, where $2/n$ is the usual multiplier for the variance of the treatment effect when a covariate is not fitted and VIF is the variance inflation factor due to non-orthogonality and is given by:

$$VIF = \frac{\sum_{i=1}^{2}\sum_{j=1}^{n} x_{ij}^2}{\sum_{i=1}^{2}\sum_{j=1}^{n} x_{ij}^2 - \frac{n}{2}\left(\bar{x}_{2.} - \bar{x}_{1.}\right)^2}.$$

However, in analysis of variance terms, the numerator is the total sum of squares and the denominator, being the difference between the total sum of squares and the between sum of squares, is the within sum of squares. Thus, we may re-write this expression as

$$VIF = \frac{SS_{Within} + SS_{Between}}{SS_{Within}} = \frac{SS_{Within}}{SS_{Within}} + \frac{1}{\left(N-2\right)}\frac{SS_{Between}}{SS_{Within}/\left(N-2\right)} = 1 + \frac{1}{\left(N-2\right)}F_{1,N-2},$$

where the last term on the right is an F statistic. In general, an $F_{\nu,\omega}$ statistic has expectation $\omega/\left(\omega-2\right)$, so substituting $N-2$ for $\omega$ in the rightmost term above we get that the expected VIF is given by

$$E\left[VIF\right] = 1 + \frac{1}{N-4} = \frac{N-3}{N-4}.$$

## Approximate Variance Inflation Factor for the stratified case

The pdf of a standard Normal truncated at 0 is

$$\phi_{trunc}\left(Z=z\right)=\sqrt{\frac{2}{\pi}}e^{-\frac{z^2}{2}},z\geq 0..$$

If a standard untruncated Normal has pdf $\phi(\ )$, then the mean of such a truncated Normal is

$$\mu_{trunc}=\sqrt{\frac{2}{\pi}}\approx 0.798$$

and the variance is

$$V_{trunc}=1-\mu_{trunc}^2=1-\frac{2}{\pi}\approx 0.363.$$

The mean square between will have expectation proportional to 0.363, whereas without stratification it would have been proportional to one. However, under $H_0$, the expectation of the total sum of squares will be unchanged. Thus, there will be an increase in the expected value of the sum of squares within. However, this effect may be expected to be small and given other uncertainties we suggest ignoring it.

A further issue is that the ratio of the mean square between to that within will not, however, have an F distribution. Nevertheless, as a rough guide to the expected maximum gain from this form of stratification the factor

$$E\left[VIF_s\right]\approx 1+\frac{0.363}{N-4}$$

seems not unreasonable.

An alternative argument for the multiplier of 0.363 proceeds like this. Stratification will achieve a median balance between the two treatment groups and this will go some way to achieving mean balance. If one is sampling from a Normal distribution, the mean is the sufficient statistic. Where that is the case, the coefficient of regression of median on mean must be one. It thus follows that the covariance between the median and the mean is equal to the variance of the mean. On the other hand, the variance of the median is asymptotically equal to $\pi/2$ times the variance of the mean. Thus the asymptotic regression of mean on median is $2/\pi$ and the proportion of variation unexplained by the median is $1-2/\pi\approx 0.363$.

In small samples, the efficiency of the median is greater than the asymptotic efficiency. On the other hand, given the sequential nature of clinical trial recruitment, perfect stratification on the median will not be possible anyway. In practice it will be difficult to do better than this.

## Correlation of the stratum indicator with polynomials.

The following formula for the moments about the origin of a random variable $X_t$ having the truncated Normal distribution is inspired by one for the standard Normal given by Rabbani[39] and uses the fact that moments of even powers will be the same as for the untruncated Normal.

$$E\left(X_t^m\right)=\begin{cases} 2^{\frac{m}{2}}\left(\dfrac{m-1}{2}\right)!\dfrac{1}{\sqrt{\pi}} & m \text{ odd} \\ 2^{-m/2}\dfrac{m!}{\left(m/2\right)!} & m \text{ even} \end{cases} \tag{11}$$

To calculate correlations, we shall also need moments about the mean: covariances and variances. However, calculation of this is simplified by the fact that the stratum indicator $S$ has mean zero and variance 1 and that the mean of any odd power of $X$ is also zero. The variance of an odd power will involve an even power, since these powers must be squared to calculate it. However, all these quantities can be given by moments of the truncated standard Normal. The formula for the correlation is then given by

$$\rho\left(m\right)=\begin{cases} \dfrac{E\left(X_t^m\right)}{\sqrt{E\left(X_t^{2m}\right)}} & m \text{ odd} \\ 0 & m \text{ even} \end{cases}, \tag{12}$$

where the moments in this formula are as given by (11).

The variance of $X^m$ and correlation of the centred stratum indicator $S$ and $X^m$ are given by $Var\left(X^m\right)=E\left(X^{2m}\right)=\left(2m-1\right)!!$ and $Var\left(S\right)=1$ we obtain for odd $m$

$$Cor\left(S,X^m\right)=\frac{2\int_0^{\infty}x^m\phi\left(x\right)dx}{\sqrt{Var\left(X^m\right)}}=\frac{2^{m/2}\left(\dfrac{m-1}{2}\right)!}{\sqrt{\pi\left(2m-1\right)!!}}=\frac{\sqrt{2}\left(m-1\right)!}{\sqrt{\pi\left(2m-1\right)!!}}. \tag{13}$$

Here, “!!” indicates the double factorial. (See, for Example, Zwillinger and Kokoska[40] section 18.7.) For even $m$ the correlation is 0. Thus, the correlations are decreasing in $m$ for $m$ odd. For example, for $m=1,3,5$ 5 the correlations are

$$\sqrt{2/\pi}\approx 0.80,\ 2\sqrt{2/\left(15\pi\right)}\approx 0.41,\text{ and }\frac{8}{3}\sqrt{\frac{2}{105\pi}}=0.21.$$

In fact, as noted before for each successive odd power the correlation coefficient is approximately halved. To obtain the partial correlations, we use the recursive formula,

$$\rho_{XY.\mathbf{W}}=\frac{\rho_{XY.\mathbf{W}\setminus(W_0)}-\rho_{XW_0.\mathbf{W}\setminus(W_0)}\rho_{W_0Y.\mathbf{W}\setminus(W_0)}}{\sqrt{1-\rho^2_{XW_0.\mathbf{W}\setminus(W_0)}}\sqrt{1-\rho^2_{W_0Y.\mathbf{W}\setminus(W_0)}}} \tag{14}$$

where $W_0\in\mathbf{W}$ and $\mathbf{W}$ is a set of covariates. The partial correlations of $X^3,X^5$ given are

$$-\frac{1}{\sqrt{3(\pi-2)}}\approx -0.54,\ -\frac{7}{6\sqrt{10(\pi-2)}}\approx -0.35.$$

The partial correlation of $X^5$ given $X^3$ and $X$ is

$$\frac{3}{2\sqrt{5(3\pi-7)}} \approx 0.43.$$